\documentclass[aps,prl,twocolumn,showpacs,groupedaddress]{revtex4}
\usepackage[dvips]{graphicx}
\usepackage{amssymb}
\usepackage{ulem}
\begin{document}
%\preprint{}
\title{Fano factor reduction on the 0.7 structure in a ballistic one-dimensional wire}
\author{P. Roche}
\author{J. S\'egala}
\author{D. C. Glattli}
\altaffiliation[Also at ]{LPA, Ecole Normale Sup\'erieure, Paris.}
\affiliation{Nanoelectronic group, Service de Physique de l'Etat Condens\'e,\\
CEA Saclay, F-91191 Gif-sur-Yvette, France}
\author{J. T. Nicholls}
\affiliation{Department of Physics, Royal Holloway, University of
London, Egham, Surrey TW20 0EX, UK}
\author{M. Pepper}
\author{A. C. Graham}
\author{K. J. Thomas}
\author{M. Y. Simmons}
\altaffiliation[Now at ]{Physics, UNSW, Sydney 2052, Australia.}
%\altaffiliation[Now at ]{School of Physics, University of New
% South Wales, Sydney 2052, Australia.}
\author{D. A. Ritchie}
\affiliation{Cavendish Laboratory, Madingley Road, Cambridge CB3
0HE, UK}
\date{\today}
\begin{abstract}
We have measured the non-equilibrium current noise in a ballistic
one-dimensional wire which exhibits an additional conductance
plateau at $0.7\times2e^2/h$. The Fano factor shows a clear
reduction on the 0.7 structure, and eventually vanishes upon
applying a strong parallel magnetic field. These results provide
experimental evidence that the 0.7 structure is associated with
two conduction channels which have different transmission
probabilities.
\end{abstract}

 \maketitle

Starting from a high mobility two-dimensional electron gas (2DEG)
the fabrication of split-gate devices have allowed the study of
one-dimensional (1D) ballistic transport. By applying a voltage
$V_g$ to the split-gate it is possible to control the number of
transverse modes transmitted through the 1D constriction created
by the split-gate, and in wires typically shorter than $1~{\rm \mu
m}$ the differential conductance characteristics $G(V_g)=dI/dV$
exhibit\cite{VanWees88PRL60p848,Wharam88} plateaus quantized at
integer multiples of $2G_0$, with $G_0=e^2/h$. The factor of two
arises from the spin degeneracy of the 1D subbands in the
constriction.

In addition to the quantized conductance plateaus, an unexpected
structure is observed near $0.7\times2G_0$. This feature, called
the 0.7 structure, appeared but was not recognized in early work
and subsequent investigations\cite{Thomas96PRL77p135} of the
effect revealed its fundamental connection with electron spin.
Although extensively
studied\cite{Thomas96PRL77p135,Thomas98PRB58p4846,
Kristensen00PRB62p10950,Reilly01PRB,Meir02PRL89n196802,
Cronenwett02PRL88n226805}, one main question remains: \textit{Does
the 0.7 structure correspond to a perfectly transmitted channel?}
Here, we give a clear experimental answer to this question by
measuring the Fano factor, $F$, of the partition noise of the
current.

For conductances $G \leq 2G_0$ there are two conducting subbands
and to understand their role in the 0.7 structure it is easiest to
consider them in a strong magnetic parallel field $B$, when they
are spin-split ($\uparrow$ and $\downarrow$) and separated by the
Zeeman energy; the conductance characteristics $G(V_g)$ show
plateaus at $G_0$ and $2G_0$. As $B$ is reduced, measurements show
\cite{Thomas96PRL77p135} that the subband separation reduces
linearly, and as $B\rightarrow0$ there remains a finite splitting.
This finite splitting could be interpreted as a simple zero-field
spin splitting, except that: (a) the lowest plateau is at $0.7
\times 2G_0=1.4 G_0$ rather than at $G_0$, and, (b) as the
temperature is lowered the 0.7 structure weakens. Measurements of
an enhanced $g$-factor as the 1D subbands are depopulated,
suggests the importance of exchange interactions
\cite{Thomas96PRL77p135}. Early calculations \cite{wang96} show
that spin splitting of the subbands is possible, and a later
phenomenological model\cite{bruus2001} based on the effects of a
dynamical local polarization in the constriction have had some
success in modelling the 0.7 structure, especially the unusual
temperature dependence. More recent microscopic
mechanisms\cite{Meir02PRL89n196802,Cronenwett02PRL88n226805,Matveev04}
are also based on the spin degree of freedom. In many theoretical
descriptions, the 0.7 structure is compatible with there being two
conduction channels with different transmission probabilities. To
date, there is no definitive proof that this is the case.

Previous noise measurements
\cite{Liu98Nature391p263,Kim03CondMat0311435} in the vicinity of
the 0.7 structure have included thermal noise contributions due to
conductance non-linearities, as well as $1/f$ noise. Furthermore,
the explored energy range (several meV) have exceeded the energy
scale of the 0.7 structure \cite{Thomas98PRB58p4846}. Here, we
present noise measurements at sub-Kelvin temperatures, with a wide
frequency range that allows us to separate the white noise and
$1/f$ noise. Moreover, a careful analysis of the non-linearities,
which are intrinsic to the 0.7 anomaly
\cite{Thomas98PRB58p4846,Cronenwett02PRL88n226805} , allows us to
extract the thermal noise variations and to obtain the pure
partition noise contribution. The deduced Fano factor shows a
reduction on the 0.7 structure. In addition, we have measured the
evolution of the Fano factor with a parallel magnetic field $B$;
at high $B$ the 0.7 structure moves to $G_0$ and the Fano factor
goes to zero.

\begin{figure}[h]
\includegraphics[angle=-90,width=7cm,keepaspectratio,clip]{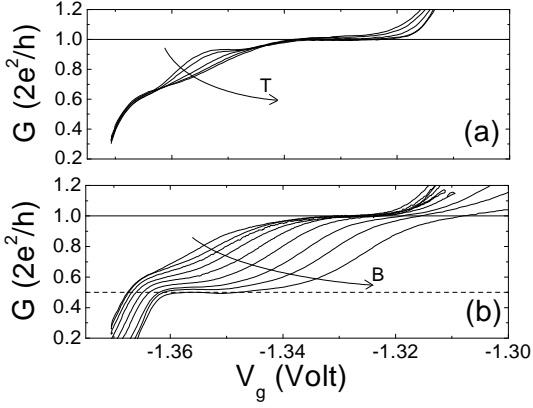}\caption{a)
Conductance characteristics $G(V_g)$ at $T=273, 360, 515, 666 $
and 779~mK. With increasing temperature the resonance at
$0.9\times2G_0$ disappears, whereas the 0.7 structure (in this
case close to $0.65 \times 2G_0)$ becomes more pronounced. (b)
$G(V_g)$ at $T=560$~mK in parallel fields $B=0$ to 8~Tesla, in
steps of 1~Tesla. With increasing $B$, the 0.7 structure evolves
into the spin-split plateau at $G_0$.}
%\label{Equi.fig}%
\end{figure}

The conductance and noise properties of a 1D conductor are well
understood using the Landauer-B\"{u}ttiker (LB) formalism, where a
scattering matrix describes a non-interacting system connected to
reservoirs. LB theory predicts shot noise suppression for
perfectly transmitted or reflected channels
\cite{Lesovik89JETP49p594,Buttiker90PRL65p2901,Martin92PRB45p1742},
which has been observed experimentally
\cite{Reznikov95PRL75p3340,Kumar96PRL76p2778}. When conduction is
linear (energy independent transmission probabilities), the Fano
factor is easy to extract as the ratio of the excess noise $\Delta
S_I(I)$ to the current $I$. Unfortunately, this is not the case
for the 0.7 structure which exhibits intrinsic non linearities
when the 1D wire is biased at energy scale $eV$ larger than the
temperature. Before analyzing our results, we will outline how
non-linearities and finite temperature are incorporated into the
LB formalism.

One considers two reservoirs (left and right) connected by $n$
channels with energy-dependent transmission probabilities
$\tau_n(\epsilon)$. The left and right reservoirs emit electrons
at energy $\epsilon$ with probabilities
$f_l(\epsilon)=f(\epsilon+eV/2)$  and
$f_r(\epsilon)=f(\epsilon-eV/2)$, where $f(\epsilon)$ is the
Fermi-Dirac distribution. The current through the sample is
$I=\frac{e}{h}\int\sum_n\tau_n(\epsilon)\left[
f_l(\epsilon)-f_r(\epsilon) \right] d\epsilon$, and the
differential conductance $G$ is an average over $k_BT$ of the
transmission probabilities at normalized energy $\pm v$,
\begin{equation}
G = {dI\over dV}=G_0 \sum_n {1\over2}\left[
\tau_n(v)+\tau_n(-v)\right], \label{g.eq}
\end{equation}
where $v= eV/2k_BT$. For energy independent transmission
probabilities, we recover the well known expression for the
conductance, $G=G_0\sum\tau_n$.

The current noise $S_I$ consists of two terms, $S_I
=S_{I~Part}+S_{I~Therm}$, the first term $S_{I~Part}$ is due to
partitioning,
% $S_{I~Part}\sim\sum\tau_n(1-\tau_n)$ which is equal to 0 when $\tau_n=1$ or 0.
and the second term $S_{I~Therm}$ results from the thermal noise
of the current emitted by reservoirs. The two noise contributions
are \cite{Martin92PRB45p1742,Buttiker92PRB46p12485}:
\begin{eqnarray}
S_{I~Part} = 2 G_0 \coth (v)\int
\sum_n\tau_n(1-\tau_n)\left[f_l-f_r\right] d\epsilon,\nonumber\\
S_{I~Therm} = 2 G_0
\int\sum_n\tau_n^2\left[f_l(1-f_l)+f_r(1-f_r)\right]d\epsilon.\nonumber
\end{eqnarray}
For energy independent $\tau_n$, the two expressions become
\begin{eqnarray}
S_{I~Part} = 2 G_0 \coth (v)\sum_n\tau_n(1-\tau_n)\times eV,\nonumber\\
S_{I~Therm} = 2 G_0
\sum_n\tau_n^2\times\left[k_BT+k_BT\right].\nonumber
\end{eqnarray}
The excess noise: $\Delta S_I(I)=S_I(I)-S_I(0)$ identifies to
$S_{I~Part}-4k_BTG(0)$ which is proportional to the Poissonian
noise $\Delta S_I=F\times S_{I~Poiss}$ with:
\begin{eqnarray}
S_{I~Poiss} & = & 2 e I \coth (v)-4k_BTG(0),\\
F & = & \frac{\sum\tau_n(1-\tau_n)}{\sum\tau_n}\nonumber.
\end{eqnarray}
This peculiar dependence of the Fano factor $F$ with transmission
has been demonstrated
\cite{Reznikov95PRL75p3340,Kumar96PRL76p2778} in shot noise
experiments in the linear regime.

In the present case, the $\tau_n(\epsilon)$ are energy dependent.
If $F$ does not vary too strongly with energy the following
approximate expressions can be derived
\begin{eqnarray}
S_{I~Part} & = & 2 e I\coth(v)F(0) \\
S_{I~Therm} & = & 2
G_0k_BT\sum_n\left[\tau_n(v)^2+\tau_n(-v)^2\right],
\end{eqnarray}
where $F(0)$ is the Fano factor averaged over $k_BT$ around zero
energy. This holds if the explored energy scale does not exceed a
few $k_BT$. The excess noise $\Delta S_I$ is not proportional to
$S_{I~Poiss}$, but $\Delta S_I=F(0)S_{I~Poiss}+\Delta
S_{I~Therm}$, where $\Delta
S_{I~Therm}=S_{I~Therm}(I)-S_{I~Therm}(0)$. \textit{Therefore
excess noise measurements are not a direct measure of the Fano
factor $F$, but also contain thermal noise variations}. As $G$
measures the mean transmission over $+v$ and $-v$, it is not
possible to know $\Delta S_{I~Therm}$; however, we can estimate a
lower bound assuming that the $n$ different modes all have the
same transmission probability at zero bias
\begin{equation}
\Delta S_{I~Therm} \ge 4G_0k_BT\left[{\tilde G(I)}^2-{\tilde
G(0)}^2\right]/n,
\end{equation}
with ${\tilde G}=G/G_0$. We will analyze the 0.7 structure
assuming that there are two modes ($n=2$ , $G\le2G_0$), with equal
transmission probabilities at zero bias. Therefore, the corrected
noise variations $\Delta S_{I~Corr}= \Delta
S_I-2G_0k_BT\left[{\tilde G(I)}^2-{\tilde G(0)}^2\right]$ can be
fitted with
\begin{equation}
\Delta S_{I~Corr}= F^+\times S_{I~Poiss}, \label{Corr.eq}
\end{equation}
where $F^+$ is a fitting parameter. Later in this paper (see
Fig.~3) the measured linear dependence of $\Delta S_{I~Corr}$ with
$S_{I~Poiss}$ validates the above assumptions and allows us to
measure $F^+$ which, because thermal contributions have not been
fully taken into account, will be an \textit{upper} bound of the
real Fano factor $F$.

The split-gate device of length 0.4~${\rm \mu m}$ and width
0.5~${\rm \mu m}$ was fabricated over
%wafer number T425
a GaAs/GaAlAs heterostructure where the 2DEG is $3400~{\rm \AA}$
below the surface, and has a density of
$1.1\times10^{11}~\rm{cm}^{-2}$ and a mobility of $2.7 \times
10^6~\rm{cm}^2/{\rm Vs}$. The 1D constriction is biased with a
current $I$ with a ${\rm 10~M\Omega}$ resistance in series, the
other side of the sample being grounded. Four-terminal
measurements are performed, and the voltage across the sample is
amplified through two independent lines using two low-noise
preamplifiers (NF Electronics LI75A) with a total gain of $1.042
\times 10^4$. The cross-correlated voltage noise
\cite{Kumar96PRL76p2778} $S_V(I,\nu)$ is measured with $\nu$ in
the $9.1-15.5$~kHz range. By measuring the Johnson-Nyquist noise
for temperatures in the range $T=200-700$~mK, the noise accuracy
has been checked to within $1\%$. We define $V$ as the voltage
across the constriction due to current biasing, taking into
account the series resistance.

Simultaneous with the noise measurements, we measure the
differential resistance $R(V_g)=dV/dI$ at 108~Hz with a 0.1~nA rms
excitation current. Figure~1(a) shows the conductance, $G(V_g) =
1/(R(V_g)-R_S(B))$, of the sample for different temperatures; the
0.7 structure is more pronounced at higher temperatures, a
hallmark of this anomaly. A series resistance $R_S=1730~\Omega$
was used to align the first quantized plateau at $2G_0$.
Figure~1(b) shows the $G(V_g)$ characteristics at $T=550$~mK in
different parallel magnetic fields $B$; as $B$ approaches 8~Tesla
the 0.7 structure evolves into the spin-split plateau at $G_0$.
The series resistance $R_S(B)$ increases with $B$ field, and from
the Shubnikov-de Haas oscillations of the 2DEG (at $V_g=0$) we
estimate a misalignment of $4.15^\circ$ between the $B$ field and
the plane of the 2DEG. The measured perpendicular component of $B$
is consistent with the $R_S(B)$ used to align the plateaus. At the
maximum field (8 Tesla), the Landau level filling factor is $7.85$
and we believe that this will not affect our findings.

\begin{figure}[h]
\includegraphics[angle=-90,width=6.5cm,keepaspectratio,clip]{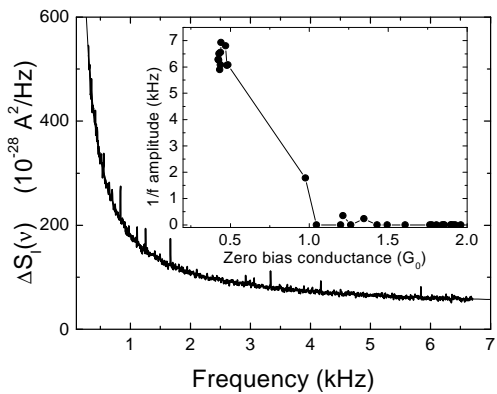}
\caption{Typical current noise power spectrum variation as a
function of the frequency for $I=10$~nA. The solid line is a fit
of white noise plus $1/f$ noise. The inset shows
the ratio $S_{1/f}(I)/S_{I~Poiss}$ as a function of the zero bias conductance.}%
%\label{Equi.fig}%
\end{figure}

Having identified the $0.7$ structure, we now focus on the noise
properties. From the measured $S_V(I,\nu)$, we deduce the current
noise power spectrum $S_I(I,\nu)=\left[1+(2\pi
RC_S\nu)^2)\right]S_V(I,\nu)/R^2$ where the shunt capacitance
$C_S=444~pF$ is measured independently.
 $S_I(I,\nu)$ contains
three distinct parts: $S_I(I,\nu)= S_{I0}(\nu)+ S_{1/f}(I)/\nu +
S_{I}(I)$. $S_{I0}(\nu)$ is the current noise applied to the
sample due to the polarization resistance and current noise of the
amplifiers, and does not depend on $R(V_g)$ or the current $I$.
The $1/f$ noise $S_{1/f}(I)/\nu$ is zero when the sample is not
current biased. $S_{I}(I)$ is the physical noise of the 1D
constriction we wish to obtain, and contains both partition noise
and thermal noise.

Figure 2 shows $\Delta S_I(I,\nu)=S_I(I,\nu)-S_I(0,\nu)$ as a
function of the frequency $\nu$, at low frequency to reveal $1/f$
noise. The frequency dependence is fitted with $\Delta
S_I(I,\nu)=S_{1/f}(I)\times\nu^\alpha + \Delta S_I(I)$,  where
$\alpha$, $S_{1/f}(I)$ and $\Delta S_I(I)$ are free parameters. In
the low frequency range, the fit is most sensitive to $\alpha$
which is found to be $\alpha=-1.007\pm0.005$; this value does not
change with current, temperature or the frequency range. We fix
$\alpha=-1$; in the measured frequency range, $9.1-15.5$~kHz, an
error of 0.005 in this exponent leads to an error of less than
0.5\% in $\Delta S_I(I)$. $S_{1/f}(I)$ is found to be roughly
proportional to $S_{I~Poiss}$ rather than to $I^2$. The Fig.~2
inset shows the ratio $S_{1/f}(I)/S_{I~Poiss}$ as a function of
the zero bias differential conductance $G(V=0)$. The amplitude of
$S_{1/f}(I)$ falls to zero near the 0.7 structure because the
transconductance $dG/dV_g$ does the same. The reduction of the
$1/f$ noise leads to an apparent noise reduction, which is not a
suppression of the partition noise.

\begin{figure}[h]
\includegraphics[angle=-90,width=7cm,keepaspectratio,clip]{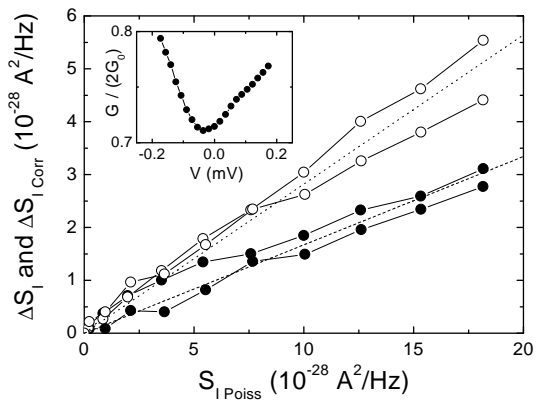}
\caption{Excess noise without ($\circ$) and with ($\bullet$)
thermal corrections, as a function of the Poissonian noise at
$T=460$~mK with $G(0)=0.71\times2G_0$. The dotted line fit to the
uncorrected data $\Delta S_{I}$ gives an overestimated $F^+=0.28$.
The dashed line fit to $\Delta S_{I~Corr}$ gives $F^+=0.17$, an
upper bound closer to $F$. The inset shows $G$ as a function of
$V$.}
%\label{Equi.fig}%
\end{figure}

With the $1/f$ noise characterized and subtracted, we determine
the Fano factor using the corrected noise variation given in
Eq.~\ref{Corr.eq}. For each determination, we measure the noise at
$\sim500$~mK by varying the bias current $I$ from $-10$~nA to
+10~nA in steps of 1~nA. The uncertainty in $\Delta S_I(I)$ is
$\sim 10^{-29}$~A$^2$/Hz. Close to pinch off, there is a
self-biasing effect due to the current, which leads to an
asymmetry in the noise plot that disappears for smaller
transconductances. The Fano factor is determined using both $+I$
and $-I$ to counter this effect.

Figure~3 shows both the raw excess noise $\Delta S_{I}$ ($\circ$)
and the corrected excess noise $\Delta S_{I~Corr}$ ($\bullet$) on
the 0.7 structure at 460~mK with $G(0)=0.71\times2G_0$. The
uncorrected measurement $\Delta S_{I}$ gives an overestimated
$F^+=0.28$, in apparent agreement with the Fano factor expected
for two channels with the same transmission probability
$(F\approx1-0.71)$. However, if thermal corrections are taken into
account, the linear variation of $\Delta S_{I~Corr}$ with
$S_{I~Poiss}$ gives $F^+=0.17$, much smaller than $0.29$. There is
a clear \textit{reduction} of the Fano factor on the 0.7 structure
\footnote{The gate voltage at which the reduction occurs does not
correspond to that of the resonance at $0.9\times2G_0$. We believe
this unwanted resonance is not responsible for the observed
reduction of the partition noise, nor for the non-linearities.}.
The accuracy on $F^+$ is $\pm0.025$, determined by both the fit
and by repetition of the measurement at fixed $V_g$. The Fig.~3
inset shows the variation of the conductance with the bias on the
0.7 structure. The observed non-linearities are similar to those
observed previously
\cite{Thomas98PRB58p4846,Cronenwett02PRL88n226805}.
\begin{figure}[h]
\includegraphics[angle=-90,width=7cm,keepaspectratio,clip]{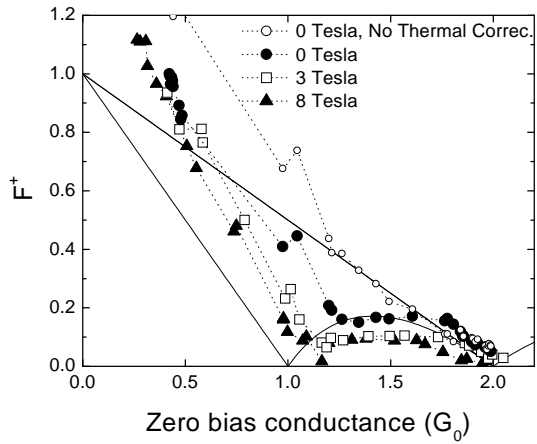}
\caption{Upper bound of the Fano factor at $T=460$~mK without
thermal corrections ($\circ$), plotted versus the zero bias
conductance. The solid circles ($\bullet$) show the same
measurements as the open circles, but with thermal corrections
applied. Squares ($\square$) show the Fano factor with thermal
corrections at $T=610$~mK and $B=3$~Tesla, and the triangles
($\blacktriangle$) are similarly corrected data at $T=580$~mK and
$B=8$~Tesla.}
%\label{Equi.fig}%
\end{figure}

Figure~4 presents the central result of our paper, where the
measured $F^+$ are plotted as a function of the zero bias
conductance. The two solid lines in Fig.~4 show the expected $F$
when there is full spin splitting ($F \rightarrow 0$ at $G_0$ and
$2G_0$) or no spin splitting ($F \rightarrow 0$ at $2G_0$). The
open circles ($\circ$) are $F^+$ obtained when thermal corrections
are not taken into account; these data do not show a reduction at
the 0.7 structure and follow the Fano factor for the case of no
spin splitting (the upper straight line). As explained previously,
because the system is non-linear these points are overestimated
upper bounds of $F$. The solid circles ($\bullet$) in Fig.~4 are
obtained from the same data as the open circles, but with the
non-linearities taken into account using Eq.~\ref{Corr.eq}; in
contrast to the uncorrected data there is a reduction close to the
0.7 structure. Whereas one cannot calculate the transmission of
the two modes from $F^+$, one can show that they are not
identical: we use thermal corrections assuming identical zero bias
transmissions of the two channels, thus $F^+$ should be above the
upper straight line of fig.4. As $F^+$ is below this line, one can
conclude that \textit{the two channels do not have the same
transmission on the 0.7 structure}. Such information can only be
obtained from simultaneous noise and conductance measurements. The
upper bound Fano factors at $B=3$ and 8~Tesla are plotted in
Fig.~4 as squares ($\square$) and triangles ($\blacktriangle$),
respectively. For both magnetic fields the suppression of the Fano
factor is more developed than for $B=0$ and the conductance at
which the reduction occurs shifts towards $G_0$. This high $B$
field result is consistent with a Zeeman splitting of the
$\uparrow$ and $\downarrow$ 1D subbands. The evolution of the
reduction with $B$ indicates that the two channels having
different transmissions at zero field may have different spin
orientations.

In conclusion, we have performed careful measurements of the Fano
factor which shows a clear reduction on the 0.7 structure. This
reduction demonstrates for the first time that the 0.7 structure
is accompanied with two conducting channels with different
transmission probabilities. The evolution of the reduction with a
parallel magnetic field $B$ supports the picture of two channels
with different spin orientations. In future it would be
interesting to measure the evolution of the Fano factor reduction
with temperature, in order to understand the underlying mechanism
which leads to these spin dependent transmissions.

%\bibliography{References}

\end{document}